\documentclass[sigconf, nonacm]{acmart}

\usepackage{times,amsmath,epsfig}
\usepackage{epstopdf}

\usepackage[ruled,vlined]{algorithm2e}
\usepackage{subfigure}

\usepackage{amsfonts}
\usepackage{graphicx}
\usepackage{diagbox}
\usepackage{multirow}
\usepackage{makecell}
\usepackage{array}

\begin{document}
\title{Proof-of-Data: A Consensus Protocol for Collaborative Intelligence}

%%
%% The "author" command and its associated commands are used to define the authors and their affiliations.
\author{Huiwen Liu}
\affiliation{%
  \institution{Singapore Management University}
  \country{Singapore}
}
\email{hwliu@smu.edu.sg}

\author{Feida Zhu}
\affiliation{%
  \institution{Singapore Management University}
  \country{Singapore}
}
\email{fdzhu@smu.edu.sg}

\author{Ling Cheng}
\affiliation{%
  \institution{Singapore Management University}
  \country{Singapore}
}
\email{lingcheng.2020@phdcs.smu.edu.sg}

%%
%% The abstract is a short summary of the work to be presented in the
%% article.
%%
%% The abstract is a summary of the work to be presented in the
%% article.
\begin{abstract}
% PoD -> FedBFT
Existing research on federated learning has been focused on the setting where learning is coordinated by a centralized entity. Yet the greatest potential of future collaborative intelligence would be unleashed in a more open and democratized setting with no central entity in a dominant role, referred to as "decentralized federated learning". New challenges arise accordingly in achieving both correct model training and fair reward allocation with collective effort among all participating nodes, especially with the threat of the Byzantine node jeopardising both tasks.

In this paper, we propose a blockchain-based decentralized Byzantine fault-tolerant federated learning framework based on a novel Proof-of-Data (PoD) consensus protocol to resolve both the "trust" and "incentive" components. By decoupling model training and contribution accounting, PoD is able to enjoy not only the benefit of learning efficiency and system liveliness from asynchronous societal-scale PoW-style learning but also the finality of consensus and reward allocation from epoch-based BFT-style voting. To mitigate false reward claims by data forgery from Byzantine attacks, a privacy-aware data verification and contribution-based reward allocation mechanism is designed to complete the framework. Our evaluation results show that PoD demonstrates performance in model training close to that of the centralized counterpart while achieving trust in consensus and fairness for reward allocation with a fault tolerance ratio of 1/3.

\end{abstract}

\maketitle

%%% do not modify the following VLDB block %%
%%% VLDB block start %%%
%\pagestyle{\vldbpagestyle}
%\begingroup\small\noindent\raggedright\textbf{PVLDB Reference Format:}\\
%\vldbauthors. \vldbtitle. PVLDB, \vldbvolume(\vldbissue): \vldbpages, \vldbyear.\\
%\href{https://doi.org/\vldbdoi}{doi:\vldbdoi}
%\endgroup
%\begingroup
%\renewcommand\thefootnote{}\footnote{\noindent
%This work is licensed under the Creative Commons BY-NC-ND 4.0 International License. Visit \url{https://creativecommons.org/licenses/by-nc-nd/4.0/} to view a copy of this license. For any use beyond those covered by this license, obtain permission by emailing \href{mailto:info@vldb.org}{info@vldb.org}. Copyright is held by the owner/author(s). Publication rights licensed to the VLDB Endowment. \\
%\raggedright Proceedings of the VLDB Endowment, Vol. \vldbvolume, No. \vldbissue\ %
%ISSN 2150-8097. \\
%\href{https://doi.org/\vldbdoi}{doi:\vldbdoi} \\
%}\addtocounter{footnote}{-1}\endgroup
%%% VLDB block end %%%

%%% do not modify the following VLDB block %%
%%% VLDB block start %%%
%\ifdefempty{\vldbavailabilityurl}{}{
%\vspace{.3cm}
%\begingroup\small\noindent\raggedright\textbf{PVLDB Artifact Availability:}\\
%The source code, data, and/or other artifacts have been made available at \url{\vldbavailabilityurl}.
%\endgroup
%}
%%% VLDB block end %%%

\section{Introduction} ~\label{sec: introduction}
% Topic: collaborative intelligence, cross-entity data exchange
% Problem: privacy and security
The data economy today is becoming increasingly collaborative in nature. Take business intelligence, for example. To unleash the full potential of big data, it is essential to integrate multi-source data depicting entities from a multi-faceted and multi-modal perspective, which, not surprisingly, is not achievable by any company alone. It is mutually beneficial for companies to leverage each other's data for collective model training. On the other hand, however, privacy and security concerns have long been major roadblocks in cross-entity data exchange. 

% We use federated learning to solve the collaborative intelligence
% with privacy and security protection
Among all the approaches proposed to resolve these data silo issues, federated learning ~\cite{konevcny2018federated} has gained growing popularity due to the fact that participating training nodes can retain all their data on-premise, train models locally and exchange only model parameters to cooperatively obtain a common model better than what each can individually train, maximally protecting their data privacy and security. This is the learning environment of collaborative intelligence we will focus on in this paper. 

% Centralized setting
% Problem: several drawbacks of centralized node
Unfortunately, existing federated learning models focus mostly on settings with a central entity to coordinate all other nodes in the training process, which we refer to as the "centralized" setting.  While useful for some scenarios, we argue that the centralized setting will not be the most important and challenging collaborative intelligence mode in the future. Competing businesses will not participate if one is in a dominating position superior to the rest in the ecosystem. For all businesses to willingly contribute and collaborate sustainably, the system must be open to all and dominated by none, which we refer to as the "decentralized" setting. 

% Decentralized setting
% Permissioned entry -> open-access
While there have been research efforts on decentralized federated learning ~\cite{hegedHus2019gossip,kim2019blockchain,majeed2019flchain,2020_li2020blockchain}, the assumption is that all participating training nodes are cooperative and motivated for a common goal in good-will spirit, typical of a consortium with permissioned entry. The main task there is to maintain model consistency across various nodes in an amicable setting.  Yet real-life application settings are never that rosy. A large-scale collaborative data intelligence ecosystem open to all must accommodate participants of all kinds, including those malicious nodes which are typically referred to as \emph{Byzantine} nodes in distributed systems. Achieving correct consensus despite the existence of Byzantine nodes is called Byzantine fault tolerance. 

% Two challenges for decentralized federated learning
% 1) How to implement correct model training in the decentralized setting?
% 2) How to implement data-based reward allocation?
% Most difficult setting: with Byzantine nodes
Two main challenges arise in this decentralized setting with Byzantine fault tolerance: (I) How to collectively train a common correct model with comparable results as in centralized federated learning; and (II) How to design fair incentivization to properly reward participating training nodes for their contribution in terms of data. The most difficult part of both challenges is to handle the "Byzantine" nodes that are largely ignored in existing studies for centralized settings. It is worth noting that in our case, Byzantine nodes not only refer to malicious nodes in traditional consensus research -- those attacking the system by compromising the consensus, which is the main task of challenge I -- but also to nodes harmful to the system by scheming for unwarranted reward allocation from their data contribution, the main task of challenge II.  

To overcome these challenges, we propose a novel consensus protocol called \emph{Proof-of-Data} (i.e., PoD) to achieve a decentralized Byzantine fault-tolerant federated learning framework.  The contributions of this paper can be summarized as follows:

%The framework utilizes distributed ledger for the recordings of the global model and model update exchange. To enable the PoD, we also devise an innovative incentive mechanism based on data contribution, which can effectively reduce malicious attacks and adequately leverage the data characteristics of each node to train the model with superior generalization performance. We then discussed the scalability of FedBFT, especially the theoretical resilience and performance analysis of PoD. 

\begin{itemize}
\item
% PoD
We propose a novel consensus protocol called Proof-of-Data tailored for decentralized federated learning with Byzantine fault tolerance. PoD combines Proof-of-Work (i.e., PoW) style asynchronous consensus with epoch-based consensus locking by a PBFT-style component, integrating the best of both by endowing, on the one hand, the missing consensus finality to the practically robust yet theoretically flawed PoW consensus and lending, on the other hand, the scalability necessary for societal-scale application setting to the otherwise sound PBFT consensus. 

\item
% Data verification
We design a privacy-preserving data verification mechanism based on P4P\cite{duan2006zero}, a zero-knowledge proof protocol, to prevent participating nodes from producing inconsistent data contributions in the training process without violating their data privacy. 
%\item We first analyze the difference between the classical and emerging federated learning from the properties, and then based on the analysis, we give a formal definition of the decentralized Byzantine federated learning problem, which systematically integrates the traditional definition and decentralized properties;

\item
% reward allocation/incentive mechanism
We devise an incentive mechanism to assess and allocate rewards based on nodes' data contribution, mitigating the risk of Byzantine attacks in terms of data contribution without sacrificing model performance economically. 

%\item
% FedBFT
%We propose a decentralized Byzantine fault-tolerant federated learning framework which integrates PoD, the zero-knowledge-proof based data verification, the incentive mechanism and distributed ledger technologies, to collectively provide all of resilience, trust and incentive to the problem of decentralized Byzantine fault-tolerant federated learning. 

\item
% Evaluation
Finally, we comprehensively evaluate the performance of the framework as well as analyze the resilience, performance and governance of PoD. An analysis of the framework's anti-attacking capability is also provided. The framework is empirically validated through a wide range of experiments on both time-invariant and time-varying datasets. The results of these experiments show that our framework performs closely with Centralized Federated Learning.
\end{itemize}

The remainder of this paper is organized as follows. First, we formulate the decentralized Byzantine federated learning (i.e., DBFL) problem and present design ideas in Section~\ref{sec: algo design} to better understand our PoD consensus protocol, which is detailed in Section~\ref {sec: PoD consensus}. In section ~\ref{sec: analysis}, the resilience, performance and governance of PoD are theoretically analyzed. Section ~\ref{sec: exp} presents the experiments and evaluates the framework's performance. Section ~\ref{sec: related work} discusses related work, and Section ~\ref{sec: conclusion} concludes this paper.

\section{Problem Formulation and Design Ideas} \label{sec: algo design}

\begin{figure*}[htbp]
\centering
\includegraphics[width=0.8\textwidth]{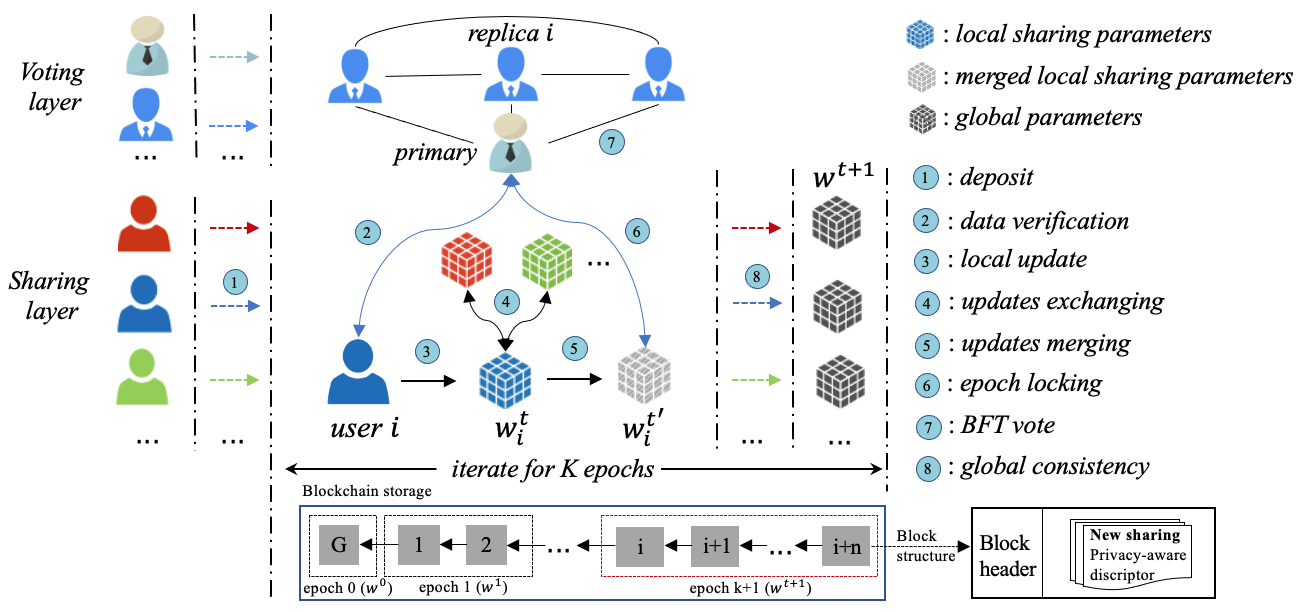}
\caption{PoD: a consensus protocol for collaborative intelligence.}
\Description{PoD consensus}
\label{fig: PoD}
\end{figure*}

\subsection{Problem Formulation} ~\label{sec: problem definition}
We follow standard notion to model the underlying environment of our problem with $n$ training devices among which several devices could be Byzantine~\cite{lamport1982byzantine} adversaries.
% Principle + graph description

%\subsection{Problem definition} ~\label{subsec: problem definition}
\emph{Definition 1. (Decentralized Byzantine Federated Learning):} Given 1) a set $\mathcal{P}$ of $n$ geo-distributed devices $\{\mathcal{P}_i\}_{i \in [n]}$ with private datasets $\{\mathcal{D}_i\}_{i \in [n]}$ connected by P2P network of which a set $\mathcal{A}$ of $f$ devices are Byzantine nodes which conduct Byzantine attacks randomly, where $[n]$ is short for $\{1,2,...,|\mathcal{P}|\}$ through the paper; 2) a model with the objective function $F(\omega)$, where $\omega$ is the model weights, the Decentralized Byzantine Federated Learning problem, denoted as DBFL, aims to make all devices $\{\mathcal{P}_i\}_{i \in [n]}$ collectively train a common model weights set $\mathcal{W} = \{\omega_i\}_{i \in [n]}$ with private datasets $\{\mathcal{D}_i\}_{i \in [n]}$ through the exchange of information over an asynchronous network.  Formally, the objective function of DBFL is to minimize the following function:
\begin{equation}
  \underset{\mathcal{W}}{\rm min\ }{G(\mathcal{W})}=\sum_{i}{p_iF(\omega_i;\mathcal{D}_i)}.
\end{equation}
Here, $\omega_i \in \mathbb{R}^d$ is the model weights of device $\mathcal{P}_i$, $p_i \ge 0$ specifies the relative impacts of each device to the whole network, and $\sum_{i}{p_i}=1$.

\subsection{Design Ideas: A Two-layer Consensus Protocol}
\label{sec: algo design}
To explain our design ideas, we start with our goal: To achieve decentralized federated learning for societal-scale applications. This context entails three essential characteristics - (1)the large number of participating nodes, (2) the existence of Byzantine nodes and (3) the absence of a central coordinating entity.  As a result, a consensus protocol is necessary to guarantee the consistency of model training and contribution accounting across different nodes with no "trust" assumption among them, lending the "trust" component to the solution. On the other hand, "incentive" component in terms of reward based on data contribution is equally indispensable to ensure the motivation of participation. The combination of both "trust" and "incentive" is the foundation of sustainable collaborative intelligence in a real-life setting.

Constrained by the FLP impossibility result~\cite{1982_FLP_impossibility}, which states that a correct consensus algorithm is impossible if three properties are to be achieved simultaneously: (I)Asynchrony, (II)Determinism and (III)Fault tolerance, the design of any consensus protocol is essentially balancing the trade-off among the three properties. 

In our setting, first of all, "fault tolerance" is indispensable as an open-access societal-scale application with neither simple faults nor Byzantine faults is simply unimaginable. "Determinism" is also deemed important because the consistency and the assurance of both model training result and reward distribution are crucial for continual participation of data-contributing nodes and liveliness of the system. We are left with "asynchrony" as the only option to let go. 

Suppose we relax the "asynchrony" property, can we use existing consensus protocol designed for synchronous or partially synchronous setting such as PBFT \cite{1999_PBFT, miller2016honey,Tendermint_buchman2018latest}?  The answer seems to be negative because a quadratic time complexity in the number of nodes is infeasible for large-scale applications as we aim for in our case.  Meanwhile, the nodes responsible for carrying out the protocol in PBFT are fixed, but in our setting, the nodes can join or quit at any time. More fundamentally, an asynchronous mode is much more desirable in our federated learning context as nodes do not need to wait for all others to complete training to benefit from the already partially-trained result. 

In order to still enjoy the efficiency from asynchrony while keeping both fault tolerance and determinism, we draw inspiration from the design of PoW (i.e., Proof-of-Work~\cite{2008_Bitcoin_pow}) as used in Bitcoin. Asynchrony (i.e., a node does not need to wait for any other node to proceed to mine independently) has played a critical role in the success of Bitcoin as the first application of consensus protocol in a societal-scale setting (18,000 public nodes as of February 2024). Unfortunately, however, PoW does not achieve the classic definition of consensus, as the finality is never secured. Specifically, it does not achieve the classic round-by-round consensus in terms of resilience, and the probability of eventually achieving global consensus increases over time, approaching infinitely close to but never reaching one. 

To remedy the situation, we propose the idea of \textbf{decoupling of model training and contribution accounting} based on the following observation:
\begin{itemize}
    \item Model training is the task performed by all the nodes most of the time with each data update. Contribution accounting, on the other hand, can be executed periodically at model training milestones when actual reward distribution is conducted. 
    \item The task to benefit the most from asynchronous processing is the model training part -- a globally consistent model training consensus is not necessary for nodes to benefit from partially trained results.
    \item The task that indeed requires finality for global consensus yet is to be performed periodically is the contribution accounting part -- a synchronous or partially synchronous protocol is possible as we do not necessarily need all the nodes to participate. 
\end{itemize}

As shown in Fig.~\ref{fig: PoD}, we, therefore, propose a two-layer architecture for the decoupling design idea, with an underlying blockchain structure to immutably record the training result and contribution accounting. The sharing layer is responsible for the asynchronous model training and generating new blocks, while the voting layer is responsible for the locking of the training result and contribution periodically (i.e., by epoch) to secure consensus finality in a partially synchronous BFT voting manner.

The benefit of decoupling the two tasks is demonstrated from our experiments (see section ~\ref{sec: exp}), where the superiority of our solution over the centralized one can be seen in that nodes can already benefit from other nodes' data by adopting local partial results due to asynchronous model training, while each node, after submitting the parameters to the central server, would have to wait for the central server to finish processing the submission from all nodes before getting back the updated parameters to use. 
\section{Proof-of-Data Consensus Protocol} \label{sec: PoD consensus}
% Two challenges for decentralized Byzantine federated learning problem
% 1) How to implement correct model training in the decentralized setting?
% 2) How to implement data-based reward allocation?
% Most difficult setting: with Byzantine nodes
% Consensus: FLP impossibility result (limitation)
The two-layer consensus protocol as introduced in Section \ref{sec: algo design}
is termed Proof-of-Data (i.e., PoD). In a DBFL system, all nodes $\{\mathcal{P}_i\}_{i \in [n]}$ join the network at random and start training the model $F(\boldsymbol{w}_i^j;\mathcal{D}_i) \rightarrow \boldsymbol{w}_i^{j+1}$ with their private datasets $\{\mathcal{D}_i\}_{i \in [n]}$, and can only exchange information employing two-party messages with P2P network with no central server to integrate information or manage distributed nodes. During the process of model training, nodes in the system generate a different sequence of model weights states $\{w_i^{j+1}\}_{i \in [n], j \geq 1}$, and some of these states may be spurious due to Byzantine nodes. Therefore, PoD takes trained model weights sequences $\{w_i^{j+1}\}_{i \in [n], j \geq 1}$ from nodes $\{\mathcal{P}_i\}_{i \in [n]}$ as inputs and aims to output a common model weights set $\mathcal{W} = w^{j+1}$. PoD guarantees the properties below except with negligible probability under the influence of any Byzantine attacks:

\begin{itemize}
\item {\verb|Safety|}: 
1) Agreement: if any two honest training nodes output $\boldsymbol{w}$ and $\boldsymbol{w'}$ for ids respectively, then $\boldsymbol{w}=\boldsymbol{w'}$;
2) Validity: if a training node outputs a model weight $\boldsymbol{w}$ for id, then it is an honest node;
3) Finality: if a model weight $\boldsymbol{w}$ is locked by voting nodes, it can not be revised any more.

\item {\verb|Liveness|}:
1) Termination: if every honest node $\mathcal{P}_i \in (\mathcal{P}-\mathcal{A})$ is activated on identification id, with taking as input a dataset $\mathcal{D}_i$ s.t. $F(\boldsymbol{w}_i^j;\mathcal{D}_i) \rightarrow \boldsymbol{w}_i^{j+1}$, then every honest node output a model weights $\boldsymbol{w}$ for id;
2) Liveliness: if a training node outputs a model weight $\boldsymbol{w^j}$ for id, then it is available to output $\boldsymbol{w^{j+1}}$ continuously.
\end{itemize}

Note that the algorithm need not reveal which nodes are faulty and that the outputs of faulty nodes may be arbitrary; it matters only that the non-faulty devices compute the same valid model weights vector for any given faulty node. Eventually, the non-faulty nodes come to a consistent view of the values of the model weight vector held by all the nodes, including the faulty ones. Fig. ~\ref{fig: PoD} shows the overview of the PoD consensus, and we will introduce three important components: block structure, sharing layer and voting layer in subsections~\ref{subsec: block structure}, ~\ref{subsec: sharing layer}, and~\ref{subsec: voting layer} respectively with two key proposed mechanisms: data verification in subsection~\ref{subsec: data verification} and measurement of data contribution in subsection~\ref{subsec: data contribution}.

\subsection{Block Structure}
\label{subsec: block structure}
%To enable authority control of the collaborative result, the storage of PoD is an encrypted blockchain system, and only authorized devices (i.e., after deposit) can access the shared contents. 

%On the blockchain, we first design blocks to store the local updates from the sharing nodes and the merging results integrating all the updates. In order to absolutely finalize the merging results on the blockchain to prevent the double-spending attacks, we design the epoch to lock the merging results. Once the epoch is locked, the information stored on the blockchain cannot be changed anymore.

%We design a detailed block structure of the PoD consensus, and the important thing to know is that the proof of data protocol is been generalized in that there is a sharing structure that can be applied to different application settings as long as this is a privacy-aware data descriptor.

The block structure in PoD would contain most of the information in a typical blockchain structure (e.g., a block header and a block body). We would focus on the information unique to PoD. 

\emph{Definition 3. (Update):} We define the training result of a device $\mathcal{P}_i$ with private dataset $\mathcal{D}_i$ as an update $U_i^h$, where $h$ is the block height. An update mainly consists of the following fields:
\begin{itemize}
\item {$w$}: the latest model weights;
\item {$ID\_Sender$}: unique identity of the sender;
\item {$data\_Summary$}: a summary that summarizes the characteristics of the data; In our work, we use the Gaussian Mixture Model (i.e., GMM~\cite{bond2001gmm}) to fit the private datasets of users and $data\_Summary$ mainly includes mean $\mu$, covariance $\Sigma$ and coefficient $\alpha$.
\item {$data\_Signature$}: a data falsification verification proving the reliability of the training results.
%\item {\verb|Validity|}: if a device outputs a model weights $\boldsymbol{w}$ for id, then id is a honest device.
\end{itemize}

In PoD, nodes will pass local updates to other nodes, who will then decide whether to merge or not. We define the merging result of local updates of sharing device $p_i^s$, $i \in [n]$ as a block $B_i^h$, where $h$ is the block height. The block body mainly stores the specific data of all the updates that have been merged in the block, and the block header consists of some specific fields related to the storage management and system settlement.

Most noticeably, an important notion \textbf{epoch} is introduced to finalize, exactly to address the finality issue in the proof of our consensus, we use epoch to lock consensus on the blockchain permanently. We denote an epoch as $E^H$, where $H$ is the epoch height. Each epoch contains a certain number of blocks (e.g.,100) with a fixed block size, and the number of blocks in an epoch depends mainly on the number of participating nodes in the system. At the end of an epoch, the voting layer must finalize the consensus by locking the current epoch. Once an epoch is locked, the information stored on the blockchain cannot be changed anymore.

\subsection{Sharing layer}
\label{subsec: sharing layer}
%In the beginning, a randomly initialized model was placed into the $0$ block, then the $0$-th round of training starts. Nodes access the current model and execute local training, and put the verified local gradients to new update blocks. When there are continuously enough update blocks, the smart contract triggers the aggregation, and a new model of the next round is generated and placed on the chain. We should note that the FL training only relies on the latest model block, and the historical block is stored for failure fallback and block verification.
Nodes in the sharing layer are responsible for handling three tasks:
 \begin{itemize}
     \item Obtain the intermediate parameters with the latest private dataset, merge updates from other devices in the network to generate a new block, and then broadcast the new block to other devices. To do that,  the node must first deposit for epoch-sharing authorization.
     \item Listen for new blocks and epoch locking to trigger blockchain replacement and block merging events.
     \item Merge the updates generated by themselves and the listened blocks from the network
 \end{itemize}

\subsubsection{Block generation}
Essentially, in centralized federated learning, the central server is mainly responsible for aggregating the parameters of all training nodes. Similarly, in the decentralized setting, we aim to lock the updates with the largest aggregative parameters for the settlement, and this idea is conceptually similar to the “longest chain.” 

In order to give priority to messages with the largest parameters, from a training node perspective, we need to do three things. First of all, we must train the model with the private dataset to generate updates. Specifically, we first deposit for epoch training authorization, then train a new update with the latest private dataset, merge updates from other nodes in the network to generate a new block, and then broadcast the new block to other nodes. At the same time, we need to constantly listen for new blocks and new epoch locking to trigger events of blockchain replacement and block merging. We need to merge the updates we generated and the listened blocks from the network. Moreover, if we receive the locking message in the middle of the training recurrence, we have to stop and merge.

\subsubsection{Block merging.} 
It is particularly important to explain why proof of work can successfully progress in an asynchronous manner. It does so because it uses the idea of the ‘longest chain’ to achieve the possibility of a global consensus in the long run. Here, we want to borrow the idea of the ‘longest chain’, but in our setting, the idea of the longest chain is the peer training node update, which contains the maximum aggregative result of the whole system’s training updates.

For each epoch, each device, including newly entered devices, starts training the model from the same initial state $w_{0}$, which is finalized in the last block of the previous epoch. After generating a new update $U_i$ for device $P_i (i \in [n])$ with the private dataset $\mathcal{D}_i$, $P_i$ packages new block and broadcasts the block $B_i^h$ to the whole network and also listens to Blocks $B$ from other devices in the network. Once a block is received from another device, the device $P_i$ integrates the received update according to the \emph{block merging protocol}. Each block in the network is the merging result of the device and has the merging list or update list. If the update list contains the received list, the block will not need to be merged. Otherwise, we will continue to merge the new block and broadcast it to the network. Meanwhile, if the block is the history block and we do not merge, we need to roll back the training process and retrain from the height of the history block. 

%Furthermore, there are two main challenges during the whole process. The first is that the local data might be fabricated, and the second is that the devices are not always available. To address the first challenge, we let the voting layer verify the actual existence of data with data summary information (see section ~\ref{subsec: data contribution}). We design a local learning progress for nodes to address the second one. Nodes can actively obtain the initial global model of the working epoch at any time and perform local training. When eligible updates are packaged on the blockchain as a reward, tokens can be attached to them at the epoch settlement.

\subsection{Voting layer}
\label{subsec: voting layer}
The finality of the consensus can not be resolved within the sharing layer. To do this, we have to overlay the voting layer on top of it, which will be a small set of nodes. The voting layer is mainly responsible for Epoch locking, data verification and value allocation. 

\subsubsection{Epoch locking}
In traditional Proof-of-X (e.g., PoW~\cite{nakamotobitcoin}, PoS~\cite{IEEEexample:shellCTANpage}), there is no deterministic finality for all transactions stored in the blockchain ledger. It's just that over time, the chance of the ledger being tampered with decreases, going to zero indefinitely, which doesn‘t mean tampering doesn't happen. However, for the PoD, tampering with the consensus result stored in the distributed ledger is easy. Therefore, we propose a BFT voting (i.e., PBFT~\cite{1999_PBFT}) layer to solve the problem. 

We divide the block height into several epochs, and at the end of the epoch, we will lock the epoch through the BFT voting. Once the epoch is locked, all the updates stored in the previous blocks can no longer be tampered with. Specifically, we maintain an active training list of the whole system. Once the primary node of the voting layer receives the signal of the epoch locking from the sharing layer, it will call for a vote according to the active training list. If more than fault tolerance threshold per cent of nodes, the voting devices pass the validation and sign for the new block containing the locking signal, then the epoch will be locked, and all other training devices will synchronize the locking epoch and start new training based on the new deterministic global model and all the updates in the previous blocks cannot be tampered any more. 

In addition, before broadcasting the signed epoch, the voting devices will make an epoch settlement to distribute rewards according to the data contribution (see section ~\ref{subsec: data contribution}). Moreover, we also listen for the data verification to prevent data forgery with privacy protection based on the P4P method (see section ~\ref{subsec: data verification}).

\subsection{Data Verification} \label{subsec: data verification}
%One important task of the voting level is data verification. If a device violates a rule, we can detect the violation and know which device violated the rule. Accountability allows us to penalize malfeasant devices. Specifically, We use the P4P~\cite{youdao2010p4p} verification mechanism to validate the existence of a private dataset for each device. Once training a result $w$, the device $\mathcal{P}_i$ actively initiates data falsification verification to the voting devices as shown in graph~\ref{Verification}. If verified, a signature $data\_Signature$ is obtained and will be encapsulated into the \emph{update} $U_i^h$. The goal of the main P4P protocal~\cite{duan2006zero} is to compute the sum of all user vectors. But it is a centralized verification structure. In our work, specifically, we map the primary node in the PBFT to serve as the central server to compute the sum of several data list to verify the existence of the private dataset combining the data summary information.

One important task of the voting level is data verification. 
If a training device violates a rule, the voting layer should detect the violation and know which device violated the rule. 
Accountability allows us to penalize malfeasant devices. 
For example, a training device has private data $D$=$\{d_1, d_2, \cdots, d_n\}$,
and claims the Gaussian distributions $G$=$\{g_1, g_2, \cdots, g_q\}$ for all features $F$=$\{f_1, f_2, \cdots, f_q\}$ in its data.
Our data verification objective is to check whether the training devices have data that match the distributions they claim.
Meanwhile, we can access devices' private data.

Concretely, we need to verify the matching for all features. 
Then, the system splits $f_j$’s value range($[Min(f_j), Max(f_j)]$) into some intervals, and the server calculates each interval's average value according to the corresponding Gaussian distribution $g_j$.
If the user has data that matches the Gaussian distribution, 
then the same interval's average sampled by the user should be similar to the average calculated by the server.
To do this, our data verification consists of two parts, namely \emph{Consistency Check} and \emph{Distribution Matching Check}

\begin{figure}[t]
\centering
\includegraphics[width=0.48\textwidth]{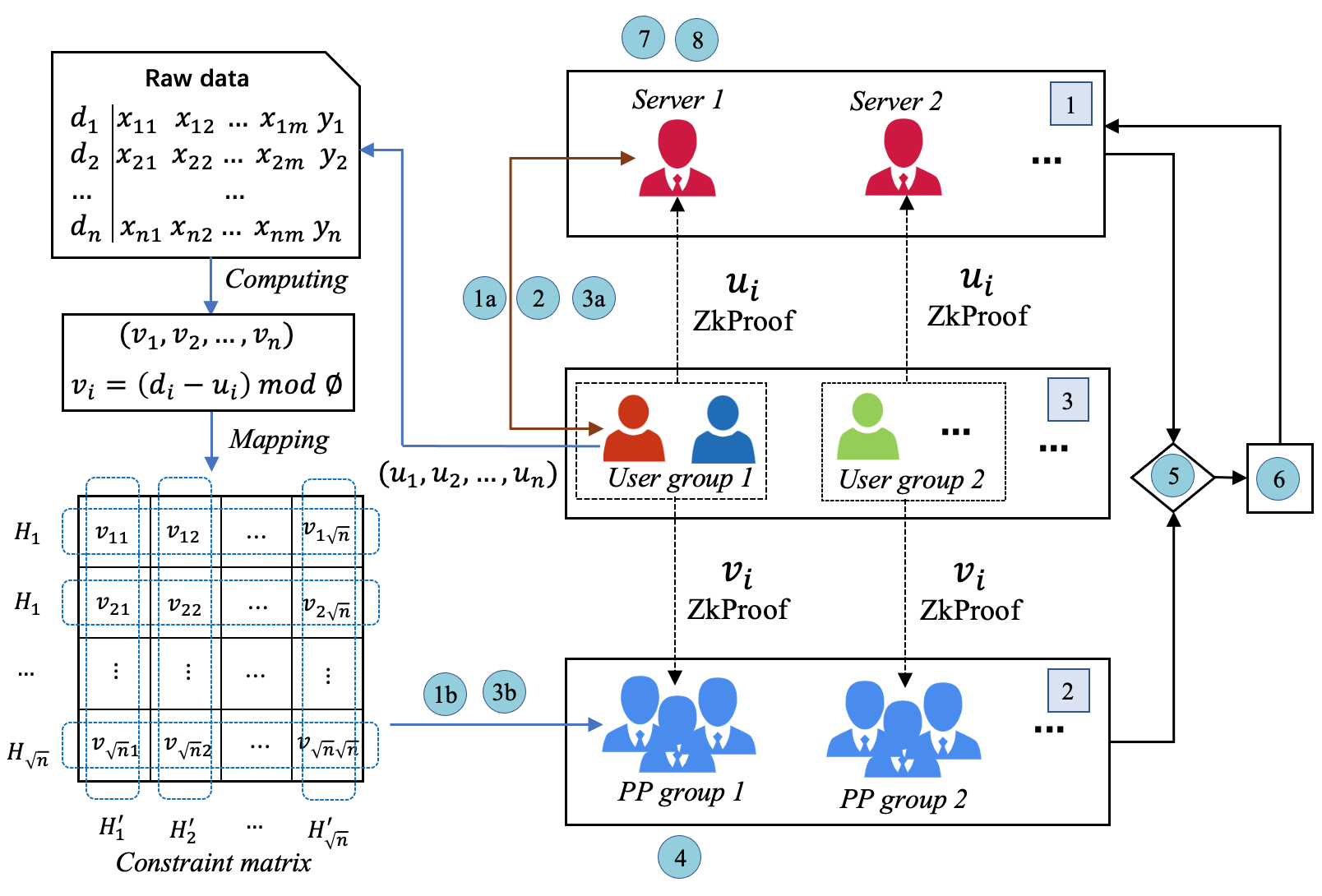}
\caption{Structure of data verification with privacy protection.}
\Description{data verification}
\label{pic: data verification}
\end{figure}

\subsubsection{Consistency Check}
To validate the 'data-distribution matching' without accessing data, 
we need three independent entities: the training device, privacy peer, and server. The training device has to send data-related information to both privacy peer and server. To validate the existence of private data and verify its consistency with server and privacy peer, we launch the P4P~\cite{youdao2010p4p} for each training device.

The system requires the user to fetch some data points in each interval and transmit their $u$ and $v$ to the system.
The system verifies the consistency of the data transmitted by the user, as shown in Algo.~\ref{algo: User_Consistency_Check}.
Under the ZK-Proof scheme, the system will broadcast several challenge vectors for corresponding independent checks.
We need the verifications for all challenge vectors to be successful.
For each independent check, the training device calculates the corresponding $u_i$ and $v_i$ for datum $d_i$ where $i\in\{1, n\}$ and send them to sever and peer respectively.
Only if all Pair-Consist-Checks are successful is this independent check verified.

\begin{algorithm}  
  \caption{Training node Consistency Check}
  \label{algo: User_Consistency_Check}
  \DontPrintSemicolon
  \SetKwInOut{Input}{input}
  \SetKwInOut{Output}{output}

  \Input{Datum $d_i$; system parameter $\phi$; system challenge vectors $C$.}
  \Output{Consistency check result $Pass$-$Flag$}
  \nl $Pass$-$Flag$ $\gets True$; \;

  \nl \For{$c_k \in C$}{
    \nl  $u_i \gets$ Generate Random Vector as $d_i$;\;
    \nl  $v_i \gets (d_i-u_i) \bmod \phi$;\;
    \nl  $CM_k \gets X_k, Y_k, S_k, B_k, Z_k \gets \mbox{Commit}(u_i, v_i, c_k, \phi)$;\;
    \nl  $Send(CM_k) \rightarrow Server, Peer$;\;
    \nl  $Pass$-$Flag$ $\mathrel{*}=$ Pair-Consist-Check(Server, Peer, $CM_k$);\;
    \nl  $Pass$-$Flag$ $\mathrel{*}=$ Pair-Consist-Check(User, Server, $X_k$);\;
    \nl  $Pass$-$Flag$ $\mathrel{*}=$ Pair-Consist-Check(User, Peer, $Y_k$);\;

    \nl \If{Not $Pass$-$Flag$}{
        \nl Break;\;
        }
    }
  \nl \Return{$Pass$-$Flag$};\;
\end{algorithm}

\subsubsection{Distribution Matching Check}
To verify whether $D$ matches $G$ without data breaching, take Feature $f_j$ as an example. First, the user gives the Gaussian distribution $g_j$ of Feature $f_j$ in her data. To prevent the user from tampering with the information provided later, the user needs to put all $v_i$ into a recording matrix (under the shape of $\sqrt{n} * \sqrt{n}$), calculate the hash value of each row and column and transmit them to the system. Therefore, the system only needs $2*\sqrt{n}$ hash values, significantly reducing the information required for verification.

The system first verifies the consistency of the data transmitted by the user, as mentioned before. After passing the \emph{Consistency Check}, the server calculates its average value through Algo.~\ref{algo: Summation Calculation}. Both Algo.~\ref{algo: User_Consistency_Check} and Algo.~\ref{algo: Summation Calculation} are same to the processes in P4P~\cite{youdao2010p4p} scheme. $F$ is a specific nonlinear function.

At the same time, for each $v$ in a specific interval, the system also requires the user to give all values of the same row and the same column in the recording matrix where $v$ is located. This is to verify whether $v$ is consistent with the original data. For $f_j$, if the verification of all intervals passes, then the verification of feature $f_j$ is successful. For a user, if the verification of all features passes, the user's data verification is successful.

\begin{algorithm}
  \caption{Training node Data Summation}
  \label{algo: Summation Calculation}
  \DontPrintSemicolon
  \SetKwInOut{Input}{input}
  \SetKwInOut{Output}{output}

  \Input{$u$ set of all data $U$; $v$ set of all data $V$; system parameter $\phi$; initial summation $A$.}
  \Output{Updated summation A'}
  \nl $\mu, \nu \gets 0$; 
  \nl $A \gets 0$; 

  \nl \For{$u_i \in U$}{
  \nl    $\mu \mathrel{+}= u_i \bmod \phi$;\;
      }
  \nl \For{$v_i \in V$}{
  \nl    $\nu \mathrel{+}= v_i \bmod \phi$;\;
      }
  \nl  $s \mathrel{+}= (\mu+\nu) \bmod \phi$;\;
  \nl  $A' = F(s, A)$;\;
  \nl \Return{$A'$}
\end{algorithm}

%Efﬁciency and Scalability: It must have adequate efﬁciency at a reasonably large scale, which is an absolute necessity for many of today’s data mining applications. The scale we are targeting is unprecedented: to support real-world applications both the number of users and the number of data items per user are assumed to be in millions. Robustness: It must be secure against realistic adversaries. Many computations either involve the participation of users or collect data from them. Cheating a small number of users is a real threat that the system must handle.

\subsection{Measurement of data contribution} \label{subsec: data contribution}
To facilitate our proposed framework, we need first to design an algorithm for the voting nodes to calculate the contribution of each user's dataset $\{\mathcal{D}_i\}_{i \in [n]}$ to the model update under the premise of privacy protection at each epoch. For this purpose, we design an algorithm to indirectly calculate the data contribution of each training node based on data summary (e.g., Gaussian fitting). For each training node, Gaussian fitting is performed locally first, and then the fitting result $G_i, i \in n$ is sent to the voting layer. At each epoch settlement, the voting layer will calculate the contribution $r_i, i \in n$ of each training node according to the Gaussian fitting results of all nodes participating in the settlement at this epoch.

% Data part: data contribution
% Model part: user score and voting score
% User score: use the local data to rank the updates from others
% voting score: using sample data

% Each training node: Gaussian fitting
\subsubsection{Sharing layer: Gaussian fitting} 
The Gaussian fitting $G_i$ for each user's private dataset is defined as a set of Gaussian functions $\{G_{ik}\}_{k \in [1,q]}$, where $q$ is the number of features after the feature extraction from raw dataset $\mathcal{D}_i$. 
%% Why do we need to extract features from the raw dataset?
Specifically, after feature extraction, features are independent, so we carry out Gaussian fitting $G_{ik}, k \in [1,q]$ for each feature, respectively. Therefore, the result of Gaussian fitting of original data $G_i$ is the combination of a series of feature Gaussian fitting result $\{G_{ik}\}_{k \in [1,q]}$. 
%The Gaussian fitting for the raw dataset is shown in Algorithm~\ref{algorithm:training_gaussian fitting}.

%\begin{algorithm}
%\caption{Training: Gaussian fitting of private dataset \label{algorithm:training_gaussian fitting}}
%\DontPrintSemicolon
%\SetKwInOut{Input}{input}
%\SetKwInOut{Output}{output}

%\Input{User private data $\mathcal{D}_i, i \in [n]$}
%\Output{Result of Gaussian fitting $G_i=\{G_{i1}, G_{i2}, ..., G_{iq}\}$, $G_{iq}$ is the Gaussian distribution of feature $q$}

%\nl $F_i$ = feature\_extraction($\mathcal{D}_i$); /*GNN*/\;
%\nl $q = $ columns of $F_i$;\;

%\nl \For {$k = 1$ to $q$} {
%\nl   $G_{ik}$ = Gaussian\_fitting($F_{ik}$);\;
%    }
    
%\nl \Return{$G_i = \{G_{i1}, G_{i2}, ..., G_{iq}\}$};\;
%\end{algorithm}

% Voting nodes: calculation of each node
\subsubsection{Voting layer: contribution calculation}
In the settlement stage, the voting layer needs to update the global model and allocate benefits according to the proportion of user data contribution. Since the voting layer only has the result of each user's Gaussian fitting result $G_i = \{G_{i1}, G_{i2},..., G_{iq}\}$ and the number of user volumes $count_i, i \in [n]$, we adopt random sampling method to approximate the real data distribution according to the all Gaussian fitting results $G^k = \{G_{mk}\}, m \in [n]$ on specific feature $k, k \in [1,q]$ to indirectly calculate the proportion of each user's data contribution $\{r_{1k},r_{2k}, ..., r_{nk}\}$ on the feature $k$. Therefore, the proportion of user data contribution $r=\{r_1, r_2,...,r_n\}$, $\sum_{i \in [1,n]}r_i=1$ and $r_i$ = $\sum_{k \in [1,q]}r_{ik}/q$.The data contribution calculation method is shown in algorithm~\ref{algorithm:voting_data_contribution}.

\begin{algorithm}
\caption{Voting: a privacy protection method for data contribution calculation \label{algorithm:voting_data_contribution}}
\DontPrintSemicolon
\SetKwInOut{Input}{input}
\SetKwInOut{Output}{output}

\Input{Gaussian fitting results set $\{G_i\}$ of all users' data $\{\mathcal{D}_i\}_{i \in [n]}$; data volumes of each user $\{count_i\}$; sample rate $r_s$; slices number $n_s$; Gaussian range $g_r$}
\Output{Percentage of all users' data contribution $r=\{r_1, r_2,...,r_n\}$}

\nl $q = $ columns of $G_i$; /*Number of features*/ \;

\nl \For {$k = 1$ to $q$} {
\nl   $G^k = \{G_{mk}\}$, $m \in [n]$;\;
\nl   $min$, $max$ = get\_max\_min($G^k$);\;
\nl   $range = (max-min)/n_s$; /*Range of each slice*/\;
\nl   \For{$i = 1$ to $n$}{
\nl      $n_s$ = $count_i * r_s$; /*Number of samples for user $i$ */\;
\nl      $p_i = \{s_j: n_j\}$ = random\_samples($G_{ik}$, $g_r$,$n_s$);\;
\nl      /*$s_j$: number of slice; */\;
\nl      /*$n_j$: number of samples in slice $s_j$;*/\; 
      }
\nl   $\{r_{1k}, r_{2k},..., r_{nk}\}$ = feature\_contribution($\{p_1, p_2, ..., p_n\}$);\;   
    }
\nl \For{$i=1$ to $n$}{
      $r_i$ = $\sum_{k \in [1,q]}r_{ik}/q$;\;
    }
\nl \Return{Proportion of all users' data contribution $r=\{r_1, r_2,...,r_n\}$};\;
\end{algorithm}

%In the following, we map our consensus algorithm to federated learning by simply changing the shared modules in the block design to model updates to form a new federated learning framework, called FedBFT. Essentially, we just replace the privacy-aware data discriptor with the training model parameters.

\subsection{An Example for flow of PoD Algorithm}
\label{subsec: pod flow}
%As we have discussed in subsection~\ref{subsec: consensus overview}, the sharing layer is mainly responsible for our asynchronous model training and generating new blocks with merging protocol, and the voting layer is mainly responsible for the partially synchronous BFT votes of epoch locking to finalize the global model. Firstly, before model training, each training node must pass the data verification to obtain the signature from the voting layer, as shown in steps 2 and 7 of Fig~\ref{fig: PoD}.
%Afterwards, the training node begins to train the model with the latest global model and merge other updates to generate new blocks, as shown in steps 3, 4, and 5. Finally, the voting layer will settle the epoch and allocate value according to each user’s data contribution to this epoch, as shown in steps 6 and 7. Importantly, we have a device management system that is responsible for token management. Devices need to deposit tokens for epoch sharing to prevent selfish attacks.

\begin{figure}[t]
\centering
\includegraphics[width=0.48\textwidth]{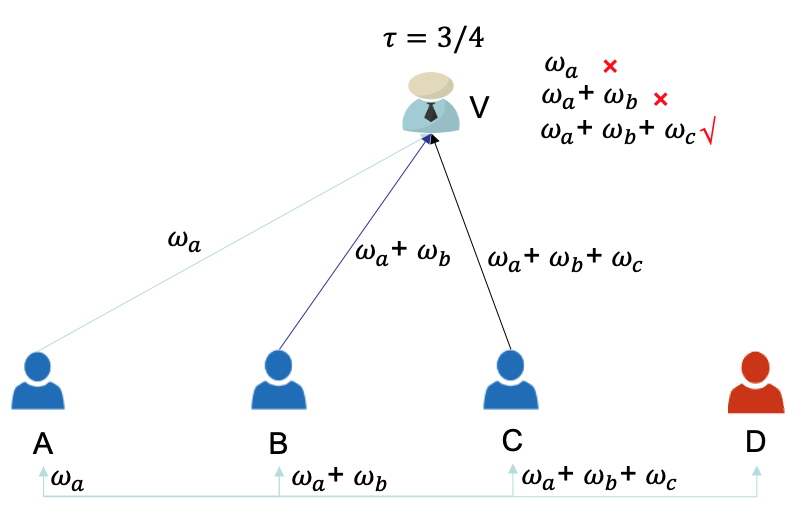}
\caption{Example of PoD flow.}
\Description{Example of PoD flow with four training nodes.}
\label{pic: PoD_flow}
\end{figure}

Supposing we have four training nodes, A, B, C, and D and one primary voting node, V, as shown in Fig.~\ref{pic: PoD_flow}.
In the process, they first train parameters locally and independently.
Suppose A is the first one to finish the training with the private dataset and gets the updated parameter vector $w_a$; then, A will broadcast the $w_a$ to training nodes B, C, and D and apply it to the voting layer for the epoch settlement. Meanwhile, A listens to the epoch-locking flag from the voting layer and new updates from B, C and D. Because the merging list only contains A’s update less than the voting threshold $\tau$, the voting layer passes the request directly. After that, if B is the second one to finish and gets the parameters update $w_b$, B will merge the $w_a$ from A with his own update, then broadcast the merged result $w_a+w_b$ to A, C and D and request to voting layer for the epoch settlement too. Similarly, this request is passed until C finishes training and submits the request with merged $w_a+w_b+w_c$. The epoch settlement will be agreed on because the merging list is more extensive than the voting threshold $\tau$. Then, the voting layer will broadcast the locking flag to all training nodes with the latest merged result $w_a+w_b+w_c$ and value settlement result. Once each training node receives the locking signal, it first updates the merging result with the latest one and then stores the result in the blockchain, and then continues to train the next $w$ continuously.

Moreover, because training node D is slow and will be left out in this epoch, the value settlement will only happen in A, B, and C. But if D catch up in the next epoch, D will also benefit from the value allocation. Moreover, suppose D is now dead or malicious in purposely holding the result; the whole system will not be affected because A, B and C will be able to proceed.

%\begin{figure}[htbp]
%\centering
%\includegraphics[width=0.48\textwidth]{pic/merging.png}
%\caption{Block merging process.}
%\label{merging}
%\end{figure}
\section{Analysis} ~\label{sec: analysis}

\begin{table*}[!h]
\tiny
\centering
    \caption{Fault tolerance of PoD}

    \begin{small}
    \begin{tabular}{|p{2.0cm}|p{4.2cm}|p{8.0cm}|}
    \hline
    \textbf{Challenges} &\textbf{Attacks} &\textbf{Defense} \\
    \hline
     \multirow{5}{*}{\makecell{Challenge \uppercase\expandafter{\romannumeral1}\\ (Model training)}}&Leave out of epoch settlement &Fault-tolerance ratio $f$; Voting threshold $\tau \leq 1-f$ \\
    \cline{2-3}
    &Delay epoch (liveliness) & Voting threshold $\tau \leq 1-f$\\
    \cline{2-3}
    &DoS attack &The priority is inversely proportional to the number of requests\\
    \cline{2-3}
    &Eclipse attack &Voting threshold $\tau \leq 1-f$ \\
    \cline{2-3}
    &Finality overturn &Finality locking at epoch\\
    \hline
    \multirow{4}{*}{\makecell{Challenge \uppercase\expandafter{\romannumeral2}\\ (Value settlement)}}&Data falsification &Data verification \\
    \cline{2-3}
    &Data domination / shadowing & Data contribution calculation tolerating overlap \\
    \cline{2-3}
    &Tamper with block data &Asynchronous encryption; Update pool to be processed for public key propagation delay\\
    \cline{2-3}
    &Value settlement overturn &Value settlement locking at epoch \\
    \hline
    \end{tabular}
    \end{small}
    \vspace{-3mm}
    \label{tb: experimental parameters}
\end{table*}

In this section, we analyze the PoD consensus on liveness, safety, performance and fault tolerance. 

% Resilience: liveness, safety and fault-tolerance
\subsection{Liveness and safety} 
% liveness: decouple model training and BFT voting
% Safety: Verification
Firstly, in the training layer, all nodes train the model with a private dataset and then broadcast the blocks to the whole network, so the timing model is asynchronous for the training layer. Moreover, the PoD uses the PBFT to realize the verification work for the voting layer, so the timing model is partially synchrony. Last but not least, PoD dictates that training nodes must wait for the verification and epoch-locking flag from the voting layer. Thus, we consider the protocol partially synchronous because of timeouts for data verification and epoch locking. So, the PoD can ensure the system's liveliness.
%Moreover, PoD is designed for the permissionless network, which is open-access to everyone including dishonest nodes who have a great probability to behaving maliciously. Therefore, PoD is Byzantine tolerant. Meanwhile, PoD leverages Gossip to broadcast information through the partially connected P2P network. 

BFT vote only happens every epoch, so it can not prevent the temporary forks of the asynchronous training layer. So the agreement is probabilistic, and the finality is temporary, but the probability increases with epoch length between two consecutive epochs due to the "longest chain". However, after epoch locking, the agreement is deterministic, and the finality is permanent. The validity is deterministic due to data verification and signature. Specifically, data verification can guarantee the correctness and consistency of raw data, and a signature from the sender's private key can block the possibility of block tampering during propagation. The termination of the PoD is deterministic because of the repeated submission of each node during the merging process and voting threshold. Repeated submission guarantees that if the node doesn't receive the epoch-locking flag from the voting network, it will keep submitting the locking requests to the voting network. Moreover, because of the partially synchronous assumption of communication between training and voting layers, the number of nodes is fixed at a certain time. Once the merging list reaches the threshold, the locking works. All of these key designs can guarantee the safety of the PoD consensus algorithm.

% Security threshold analysis
\subsection{Fault-tolerant threshold analysis} % fault-tolerance
% PBFT 1/3
% model training 
In the PoD, nodes in both layers participate in the consensus-reaching process.
The voting layer is a classic PBFT model that tolerates no more than $\lfloor m \rfloor$ faulty nodes based on the conclusion $Z \geq 3f+1$ ~\cite{2002_PBFT_proactive_recovery}.

In the sharing layer, if it is a Proof-based consensus model, we need to analyze the threshold of consensus-reached sharing nodes required to ensure the security and liveness of the whole system. During the consensus-reaching process, as there is a voting threshold $\tau$ for the voting layer to validate the epoch-locking application, the $\tau \leq 1-f$.

%Performance
\subsection{Communication complexity}
The PoD model is proposed in Fig.~\ref{fig: PoD}, where the sharing layer has $n$ sharing nodes, and the voting layer has $m$ voting nodes. In the voting layer, for peer-to-peer communication, the communication complexity is the square of the node number. Thus, the required complexity $C_v$ to reach consensus is $C_v = m^2$. Moreover, we investigate the merging and locking process with minimum communication complexity for a proof-based system $C_t = n$. But based on the design of PoD, we can get that $m$ is much less than $n$, so the communication complexity of the PoD is $n$.

% Anti-attacks
\subsection{Proof of Correctness} 
To prove the correctness of PoD, we must first prove that we have successfully addressed the two challenges which we have analyzed in the section~\ref{sec: algo design}. For challenge one, we need to prove that our model training is correct. That means our decentralized version can achieve what a centralized version can do. Then, we are going to examine possible ways that a malicious node can tamper with the system. Similarly, with the training nodes A, B, C and D as examples, as shown in Fig.~\ref{pic: PoD_flow}, suppose the D is the malicious node. What are the possible ways D can ruin the system?

Firstly, D can train fast to rush into the epoch locking and leave out all others in the system to achieve the most benefits. Our solution is that we have the Voting threshold $\tau \leq 1-f$ in the voting layer to make sure that less than fault-tolerant ratio malicious nodes will not be able to dominate the system. Secondly, D can run slow intentionally to delay epoch locking, which ruins the system's liveliness. Similarly, we have the voting threshold in the voting layer to ensure that malicious nodes with a lower fault-tolerant ratio cannot affect the settlement of the epoch. Next, D can construct multiple transactions to submit epoch settlement requests redundantly to block the primary node in the voting layer (i.e., DoS attack). D may continuously broadcast his demand to the voting layer and not perform any subsequent operations. Our solution sets a request priority, which is inversely proportional to the number of locking requests. Also, D can collude other training nodes to launch Eclipse attack, our voting threshold can protect the system from this attack. At last, depending on the epoch locking at epoch settlement, the finality can not be overturned.

For challenge two, we need to prove that our value settlement is correct. That means our value allocation is fair and cannot be overturned. Similarly, D can launch data falsification attacks, and our solution is to have data verification to ensure data consistency during model training. In order to encourage training nodes to attend the system, especially those with small datasets, our data contribution calculation can tolerate overlap to protect fairness from data domination or data shadowing. Moreover, our solution uses asynchronous encryption to protect the block data consistency from tampering. At last, depending on the epoch locking at epoch settlement, the value settlement can not be overturned.

%Governance
%\subsection{Governance.}
%For the governance of the PoD consensus, we mainly focus on Decentralization, incentive, and supervision. For decentralization, depends on PoD, the certainty is probabilistic and the coverage is full, but depends on the BFT vote, the certainty is deterministic and the coverage is partial. The cost of the PoD consensus mainly is the deposit for training authorization of the training device. Moreover, for the training device, the benefit is a token reward from data contribution. For the voting device, the benefit is the token reward from epoch voting and data verification. For the supervision, the reward is the “Finder’s fee” of the submitter of the slashing condition of the BFT vote, and the penalty is the forfeit of the deposit for malicious behaviors of training devices. Therefore, the data verification, deposit and reward allocation guarantee the fairness of the PoD, and the incentive from tokenomics and fairness guarantees the liveliness of PoD.
\section{Experiments and Evaluation} ~\label{sec: exp}
In this section, we first declare the experimental setup and then evaluate the performance and fairness of PoD on dataset ImageNet~\cite{deng2009imagenet} with varied data allocation for all training nodes.

\subsection{Experimental setup}
\subsubsection{Dataset}
The publicly accessible dataset ImageNet~\cite{deng2009imagenet}, widely used in the image classification field, is used for our performance and fairness validation in this chapter. We extracted a subset from ImageNet, which comprises about 1,500,000 images and a total of 1,000 categories, with a training set of 1,400,000 examples and a test set of 100,000 examples, respectively.

\subsubsection{Dataset allocation}
% non-independent identically distribution
% Data variation: static vs. dynamic
% Data overlap: data intersection; data with intersection and data without intersection (data intersection ratio)
% Data volumes: uniform vs. uneven
In our experiments, we aim to verify system performance and system fairness in different data distribution scenarios. We mainly simulate the practical application scenarios from three aspects, i.e., data variation (i.e., static vs. dynamic), data overlap (i.e., data intersection ratio) and data volume (i.e., small vs. large dataset). (1) Data variation: We mainly consider static and dynamic data cases and design data that increase the rate of dynamic situations. (2) Data overlap: We also consider the intersection of user data. \emph{Intersection ratio} represents data overlap, and in the previous discussion, the greater the data overlap, the lower the data contribution. (3) Data volumes: Moreover, the amount of user data can greatly influence consensus results, so we also take data volumes into account during the validation process. Specifically, We use the \emph{uneven rate} to measure the unbalanced volume distribution of user datasets.

Therefore, we designed two ways for data allocation for each data variation scenario in our experiments, i.e., data volume imbalance and overlap rate difference allocations. (1) data volume imbalance: each user is assigned a different number of training samples $\{D_i\}_{i \in [n]}$ with a uniformly random distribution over 1,000 classes, and there is no overlap of any two datasets, i.e., $D_i \cap D_j = \emptyset, 1 \leq i < j \leq n $. We use the uneven rate $r_{uneven} = 1/n \sqrt{\sum_{i \in [n]}{(count_i-avg)^2}}$, where $avg = 1/n \sum_{i \in [n]}count_i$. (2) Overlap rate difference: each user is assigned the same number of training samples with a uniformly random distribution over 1,000 classes, but there exists an overlap between any two datasets, i.e., $D_i \cap D_j \neq \emptyset, 1 \leq i < j \leq n $. We use the overlap rate $r_{overlap} = n_{overlap}/\sum_{i \in [n]}count_i$, where $n_{overlap}$ represents the data number of all intersections.

It should be emphasized here that, for practical application purposes, we only consider independent identical distribution (i.e., i.i.d) data in our experiment. Moreover, In actual applications, user data basically remains unchanged, so our experiments are mainly focused on static scenarios. 

\subsubsection{Neural network structure}
For the training models used to perform the image classification tasks, we use the convolutional neural network (CNN) for ImageNet, which contains two 5$\times$5 convolutional layers (each layer is followed with a batch normalization and 2$\times$2 max pooling), a fully connected layer with ReLu activation and a final softmax output layer. The CNN model is mended from~\cite{mcmahan2017communication}. Unless otherwise specified, some important hyperparameters in our experiments are set as Table~\ref{tb: experimental parameters}.

\begin{table}[!h]
\tiny
\centering
    \caption{Experimental parameters setting of PoD}

    \begin{small}
    \begin{tabular}{|p{1.8cm}|p{2.08cm}|p{3.5cm}|}
    \hline
    \textbf{Layer} &\textbf{Parameter} &\textbf{Numerical value} \\

    \hline
    \multirow{2}{*}{Sharing layer} &\#Training nodes &4000 \\
    \cline{2-3}
    &\#Byzantine nodes&0; 1000 (1/4); 2000 (1/2) \\
    \hline
    \multirow{2}{*}{Voting layer}&\#Voting nodes &10 \\
    \cline{2-3}
    &\#Byzantine nodes&$\leq$ 3 (PBFT fault tolerance)\\
    \hline
    
    \end{tabular}
    \end{small}
    \vspace{-3mm}
    \label{tb: experimental parameters}
\end{table}

\subsubsection{Baseline and Metric}
% Baseline: centralized federated learning
In this work, we compare our PoD with a conventional centralized federated learning method (e.g., Fedavg~\cite{nilsson2018performance}) and PBFT. For the FedAvg and PBFT implementation in this chapter, a centralized topology with the same number of training nodes as the decentralized topology is considered, where more than 80\% of all training nodes are ensured to participate in each training round.
% Metric: test accuracy; system difference, max difference
To make the result clear, we design two metrics, including \emph{Acc} and \emph{Diff}, to indicate the performances of different methods and the fairness of the proposed framework. Specifically, for performance, we respectively test the global model of each method and get the test accuracy (i.e., \emph{Acc}) of the global model after each epoch. Generally, a superior federated learning method is expected to obtain a higher \emph{Acc}. For fairness, we calculate theoretical and actual rewards for each training node respectively and get the maximum reward difference (i.e., \emph{Max-diff}) among all training nodes and the total reward difference (i.e., \emph{Sys-diff}) of the system. Generally, a fairness federated learning system is expected to obtain a smaller \emph{Max-diff} and \emph{Sys-diff}).
% Byzantine
Moreover, we consider the quantitive metric $f$-resilient to characterize the resilience PoD protocol. A consensus protocol is said to be $ f$-resilient if it can tolerate an (adaptive) adversary that corrupts up to $f$ devices. If $3f +1 = n$, the consensus protocol is said to be optimally resilient. Through the paper, we focus on the optimally resilient DFL framework against an adaptive adversary.

\subsection{Performance and fairness on static data} \label{Uneven data without overlap}
Fig.~\ref{1: data volume} and Fig.~\ref{2: data overlap} respectively show the performances and fairness of PoD with static data allocation under different data volume distribution and data overlap rates.  The hyper-parameters in this experiment follow Table~\ref{tb: experimental parameters}.

\subsubsection{Uneven data without overlap}
% with or without Byzantine nodes
% Dialectical relation discussion between threshold vs. f-Byzantine
% Figure 1: without Byzantine, Ratio overlap = 0, data volume = uniform -> uneven
% 1) Acc; 2) sys_diff; 3) max_diff
% uneven rate = distance(number of each user data, avg data)
\begin{figure*}[htbp]
	\centering
	\begin{minipage}{0.32\linewidth}
		\centering
		\includegraphics[width=0.9\linewidth]{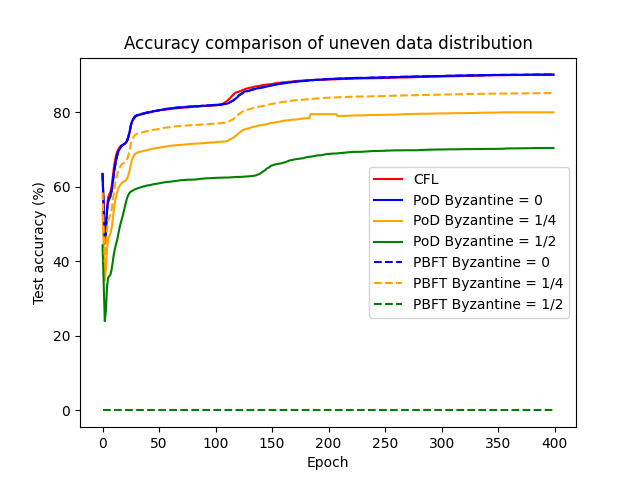}
		\centerline{(a) $r_{uneven} = 0$}
	\end{minipage}
	\begin{minipage}{0.32\linewidth}
		\centering
		\includegraphics[width=0.9\linewidth]{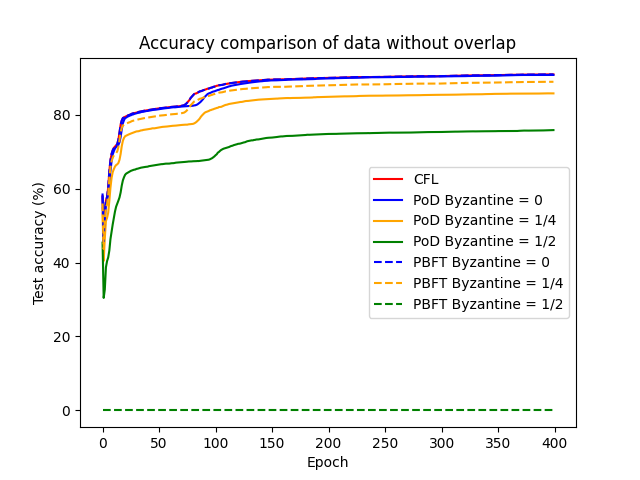}
		\centerline{(b) $r_{uneven} = 2273.03$}
	\end{minipage}
	\begin{minipage}{0.32\linewidth}
		\centering
		\includegraphics[width=0.9\linewidth]{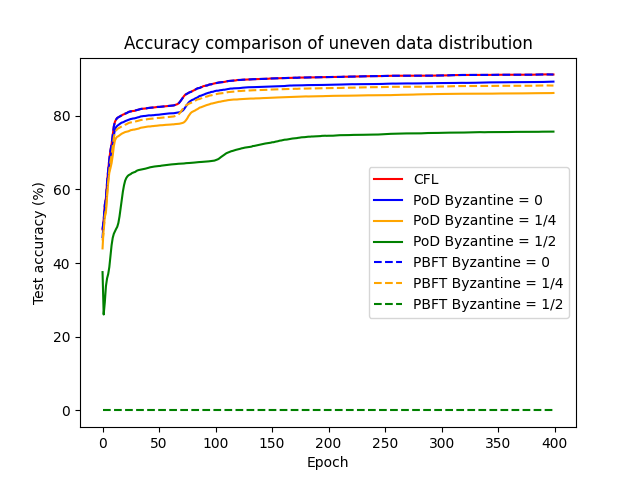}
		\centerline{(c) $r_{uneven} = 4326.92$}
	\end{minipage}
	%\qquad
	
	\begin{minipage}{0.32\linewidth}
		\centering
		\includegraphics[width=0.9\linewidth]{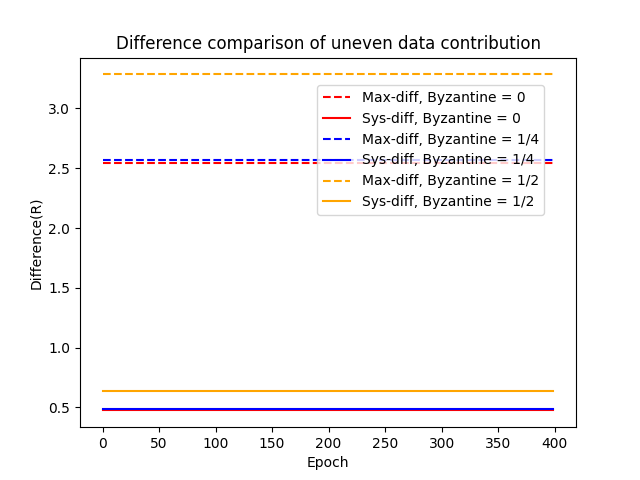}
		\centerline{(d) $r_{uneven} = 0$}
	\end{minipage}
	\begin{minipage}{0.32\linewidth}
		\centering
		\includegraphics[width=0.9\linewidth]{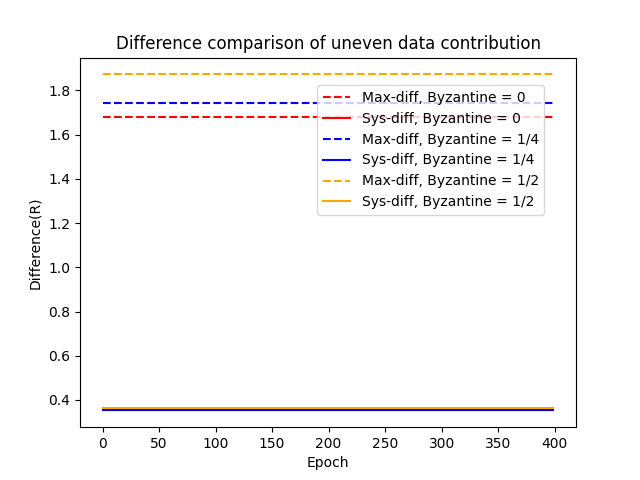}
		\centerline{(e) $r_{uneven} = 2273.03$}
	\end{minipage}
	\begin{minipage}{0.32\linewidth}
		\centering
		\includegraphics[width=0.9\linewidth]{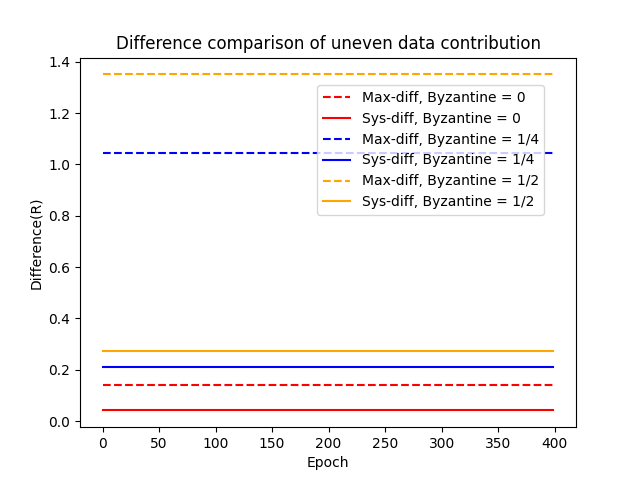}
		\centerline{(f) $r_{uneven} = 4326.92$}
	\end{minipage}
	\caption{The result of performance and fairness under different data volume distribution}
	\label{1: data volume}
\end{figure*}

Fig.~\ref{1: data volume} shows the result with static data and different data volume distributions. From the following three perspectives, we can draw different conclusions.

% performance analysis
First, the proposed PoD consensus protocol has a close performance to conventional centralized federated learning (i.e., CFL) and PBFT in terms of \emph{Acc} under almost uniform data volume distribution without Byzantine nodes, albeit slightly inferior under extremely uneven data distribution. Specifically, from Fig.~\ref{1: data volume}(a), Fig.~\ref{1: data volume}(b), the PoD finally harvests 92.02\% and 92.01\% accuracy, respectively, under uniform and slightly uneven data distribution. This is closer to the result of CFL with 92.03\% and 92.03\% accuracy and PBFT with 92.03\% and 92.03\% accuracy under the same data distribution. However, from Fig.~\ref{1: data volume}(c), PoD harvests 87.01\% accuracy under extremely uneven data distribution, which is inferior to the result of CLF with 92.02\% under the same data distribution. Since under almost uniform data distribution, all training nodes have a closer training speed. Therefore, regardless of the voting threshold, all users' updates can participate in the settlement at each epoch and be written to the blockchain. As a result, PoD training results are close to those of conventional CFL. Contrarily, extremely different training speeds exist under extreme data volume distribution, which may severely rely on the threshold of voting limits. If the threshold is very high, the waiting time of the system becomes longer, and more training time is given to the big data training node. If the threshold is very low, the big data node does not have enough time to complete training before the system epoch settlement and cannot participate in the system settlement. This is exactly why PoD underperforms conventional CFL under extreme data volume distribution in this experiment.

% fairness analysis
% << 1
Second, the PoD consensus protocol is more fair in the case of almost uniform data distribution but less fair in the case of extremely uneven data. That is, the \emph{Max-diff} is close to the \emph{Sys-diff} under the almost uniform data distribution, and both are significantly smaller than extremely uneven data distribution. Specifically, from Fig.~\ref{1: data volume}(c) and Fig.~\ref{1: data volume}(d), the PoD has 0.042 and 0.045 system differences, and the corresponding 0.071 and 0.073 maximum differences under the two almost uniform data distribution, which are close to the system differences and much less than 1, while has 1.4 system difference and 1.8 maximum difference under the extremely uneven data distribution as shown in Fig.~\ref{1: data volume}(e). On the one hand, because the PoD consensus uses Gaussian fitting to approximate the calculation of user contribution, there is a small system difference. However, since the maximum difference is similar to the system difference, the contribution measurement deviation of each user is small, and the system is relatively fair. On the other hand, as the data distribution is extremely uneven, big data users cannot participate in settlement under the pledge condition, so there will be significant inequity.

% anti-attacking analysis, from the performance and fairness
Third, the attacks from Byzantine nodes can significantly affect the system's performance and can result in significant unfairness. Rather, these Byzantine nodes will suffer profit losses. Specifically, from Fig.~\ref{1: data volume}(a), Fig.~\ref{1: data volume}(b) and Fig.~\ref{1: data volume}(c), it can be seen that the test accuracy of PoD-Byzantine is significantly inferior to that of PoD without Byzantine nodes. This slight discrepancy can be interpreted as a loss of the Byzantine nodes' own data that does not participate in the computation and the lower threshold $\tau$ that honest nodes cannot participate in the settlement, but the attacking behaviour of Byzantine nodes can significantly affect the system. Moreover, from Fig.~\ref{1: data volume}(d), Fig.~\ref{1: data volume}(e) and Fig.~\ref{1: data volume}(f), it can be seen that the system difference and max difference of PoD-Byzantine are larger than those of PoD without Byzantine nodes. In order to resist attacks of Byzantine nodes and guarantee the liveliness of the system, PoD-Byzantine sets a lower threshold $\tau$ than honest PoD that causes several honest training nodes to be unable to participate in epoch settlement, which results in significant unfairness. Moreover, the deposits of Byzantine nodes are also awarded to honest nodes participating in the epoch settlement.

In summary, the PoD consensus protocol performs closely to conventional CFL and PBFT under almost uniform data contribution and has excellent fairness without Byzantine nodes. PoD consensus protocol is Byzantine tolerance, and the tolerant ability is relevant to the voting threshold $\tau$. The fact is, there is a game between fairness (i.e., higher threshold $\tau$) and liveliness (i.e., lower threshold $\tau$) of the Byzantine system, and we need to look for a balance to maximize the performance of the system.

\subsubsection{Uniform data with overlap}
\begin{figure*}[htbp]
	\centering
	\begin{minipage}{0.32\linewidth}
		\centering
		\includegraphics[width=0.9\linewidth]{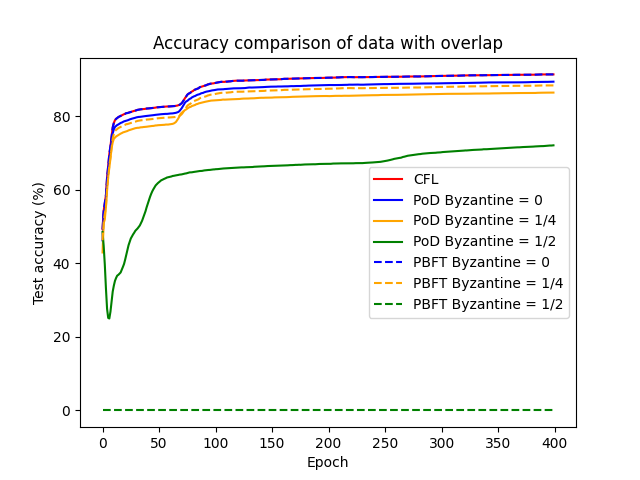}
		\centerline{(a) $r_{overlap} = 0.1$}
	\end{minipage}
	\begin{minipage}{0.32\linewidth}
		\centering
		\includegraphics[width=0.9\linewidth]{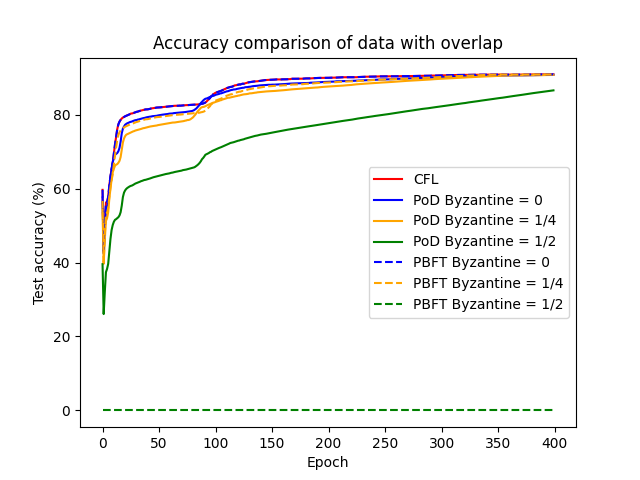}
		\centerline{(b) $r_{overlap} = 0.5$}
	\end{minipage}
	\begin{minipage}{0.32\linewidth}
		\centering
		\includegraphics[width=0.9\linewidth]{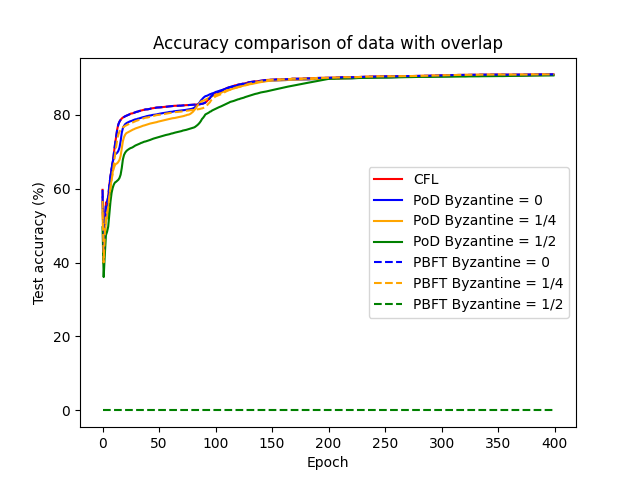}
		\centerline{(c) $r_{overlap} = 0.9$}
	\end{minipage}
	%\qquad
	
	\begin{minipage}{0.32\linewidth}
		\centering
		\includegraphics[width=0.9\linewidth]{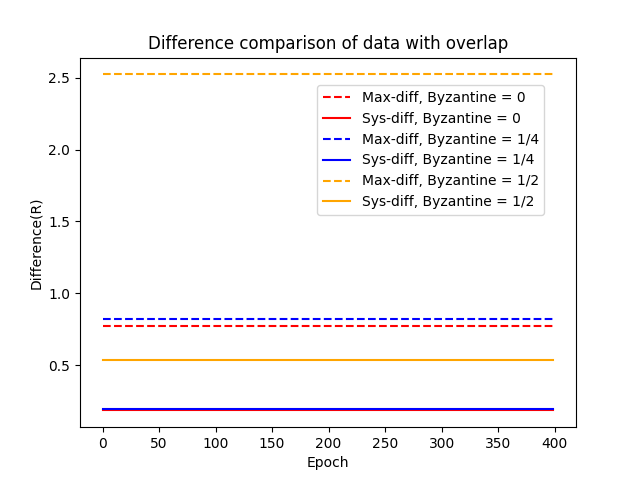}
		\centerline{(d) $r_{overlap} = 0.1$}
	\end{minipage}
	\begin{minipage}{0.32\linewidth}
		\centering
		\includegraphics[width=0.9\linewidth]{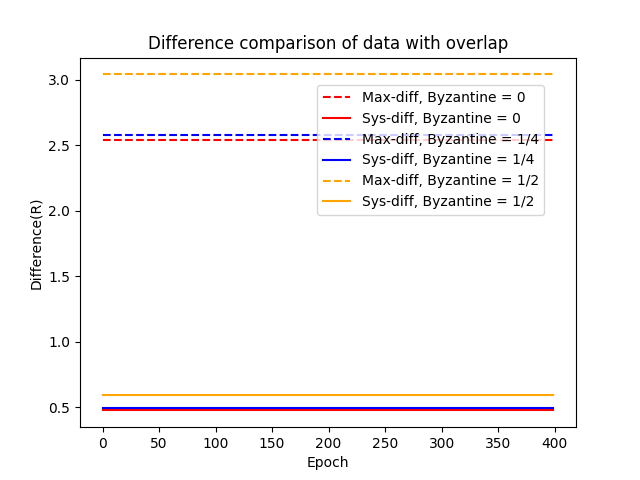}
		\centerline{(e) $r_{overlap} = 0.5$}
	\end{minipage}
	\begin{minipage}{0.32\linewidth}
		\centering
		\includegraphics[width=0.9\linewidth]{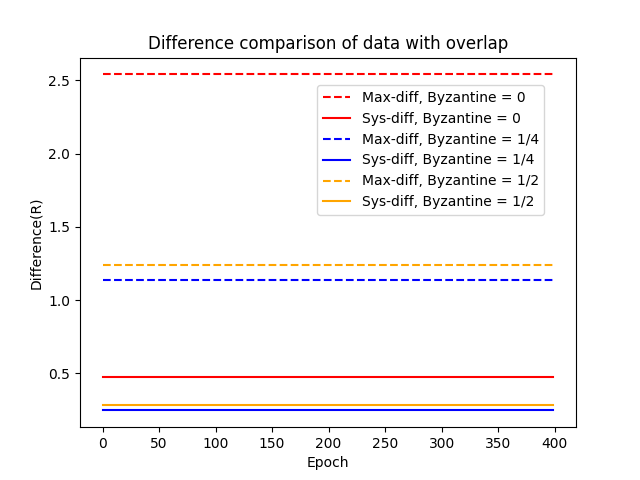}
		\centerline{(f) $r_{overlap} = 0.9$}
	\end{minipage}
	\caption{The result of performance and fairness under different overlap rate}
	\label{2: data overlap}
\end{figure*}

In this section, we investigate how the data overlap affects the performance and fairness of the PoD consensus protocol. Fig.~\ref{2: data overlap} presents the result under the uniform data with overlap.

From the perspective of \emph{Acc}, the proposed PoD consensus protocol has a close performance to the conventional CFL and PBFT under uniform data volume distribution with overlap. Take the result on overlap = 50\% as an example(Fig.~\ref{2: data overlap}(d)). The PoD finally reaches 92.03\% accuracy, which is close to the CFL with 92.07\% and PBFT with 92.02\%. Byzantine nodes can only affect the speed of convergence but cannot affect the training result. Intuitively, the PoD consensus protocol has a novel contribution measurement system to support the \emph{global aggregation} phase. User data contribution measurement system measures user aggregating factors from data volume and data uniqueness comprehensively, unlike the conventional CFL, which only considers the user data volume. Therefore, the PoD consensus can effectively eliminate the impact of data redundancy on system performance.

Besides, for the \emph{Sys-diff} and \emph{Max-diff} of honest PoD shown by Fig.~\ref{2: data overlap}(d), Fig.~\ref{2: data overlap}(e) and Fig.~\ref{2: data overlap}(f), results similar to that under uniform data volume distribution as shown in Fig.~\ref{1: data volume}(d). As shown in the figures, the honest PoD also has similar system differences and maximum differences under the uniform data distribution, and both differences are much less than 3.0. This result indicates that the system's fairness is closely related to the distribution of data volume but not to the data overlap. Finally, Pod consensus can not defend against \emph{data redundancy attack}. System performance and fairness are significantly reduced.

In summary, with the support of a user data contribution measurement system, data overlap does not significantly affect the performance and fairness of PoD. However, because the PoD consensus does not consider the quality of model updates, the system cannot resist \emph{data redundancy attack}.

\subsection{Influences of dynamic changes}
The section~\ref{Uneven data without overlap} has shown the result of PoD consensus on static datasets and voting threshold $\tau$ and declared its practicality under both data volumes and data overlap distribution. In this section, we test the performance and fairness on dynamic datasets and voting threshold $\tau$ and show the result in Fig.~\ref{3: dynamic data volume}. The hyper-parameters in this experiment follow Table~\ref{tb: experimental parameters}.

\subsubsection{Dynamic data volume without overlap}
\begin{figure*}[htbp]
	\centering
	\begin{minipage}{0.32\linewidth}
		\centering
		\includegraphics[width=0.9\linewidth]{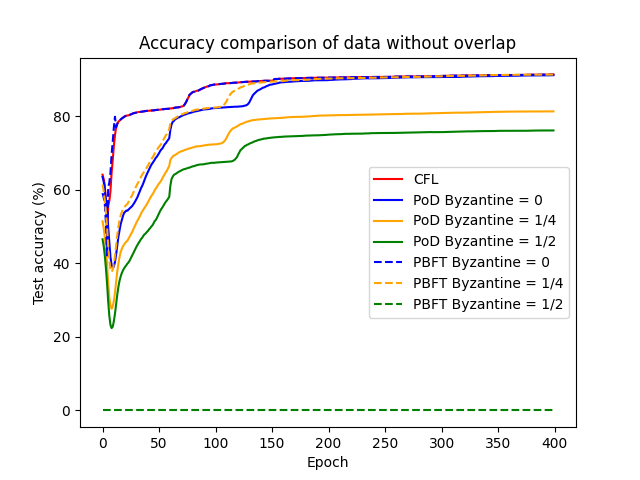}
		\centerline{(a) Dynamic data volume without overlap}
	\end{minipage}
	\begin{minipage}{0.32\linewidth}
		\centering
		\includegraphics[width=0.9\linewidth]{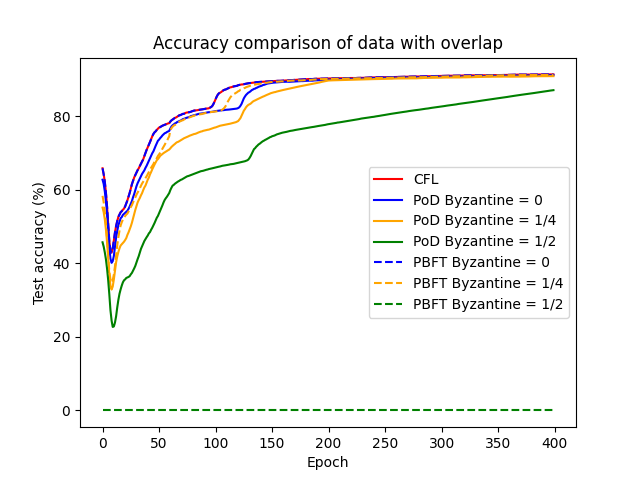}
		\centerline{(b) Dynamic data volume with overlap}
	\end{minipage}
	\begin{minipage}{0.32\linewidth}
		\centering
		\includegraphics[width=0.9\linewidth]{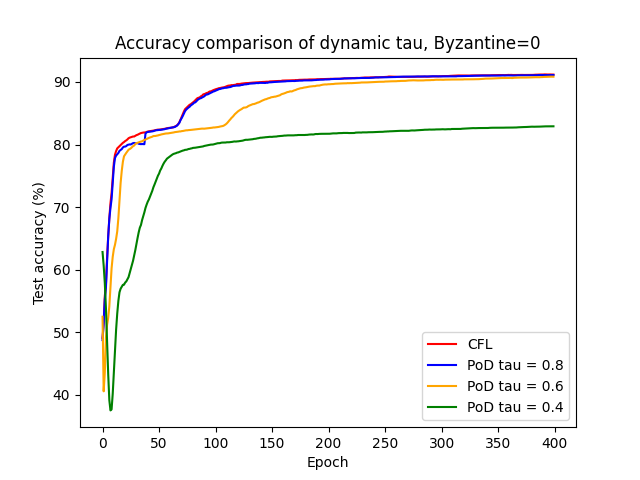}
		\centerline{(c) Dynamic threshold $\tau$}
	\end{minipage}
	%\qquad
	
	\begin{minipage}{0.32\linewidth}
		\centering
		\includegraphics[width=0.9\linewidth]{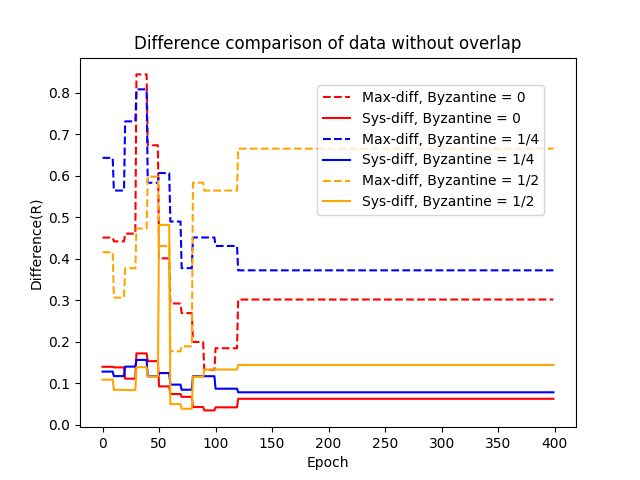}
		\centerline{(d) Dynamic data volume without overlap}
	\end{minipage}
	\begin{minipage}{0.32\linewidth}
		\centering
		\includegraphics[width=0.9\linewidth]{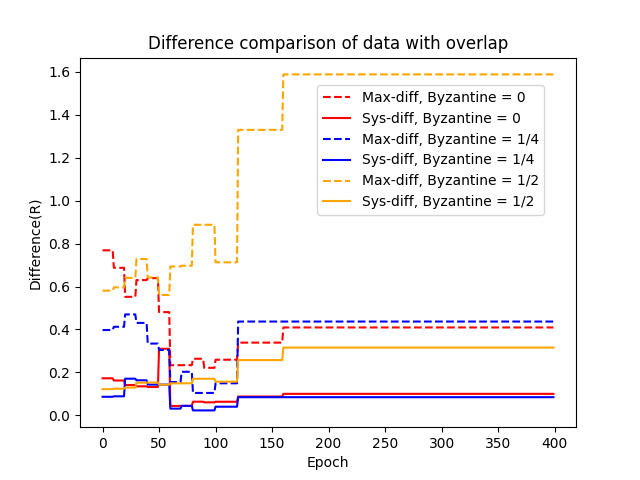}
		\centerline{(e) Dynamic data volume with overlap}
	\end{minipage}
	\begin{minipage}{0.32\linewidth}
		\centering
		\includegraphics[width=0.9\linewidth]{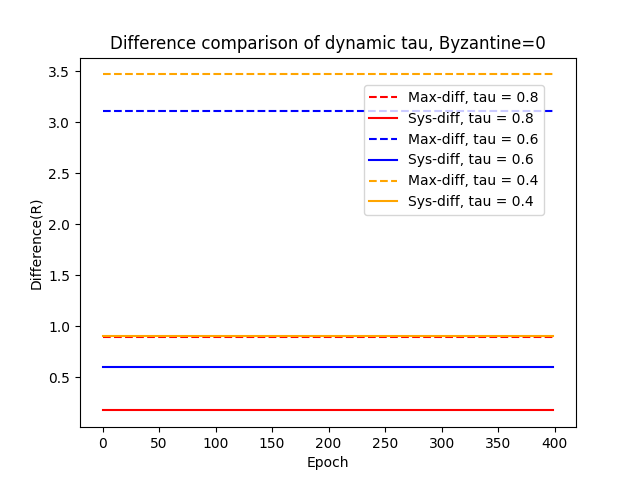}
		\centerline{(f) Dynamic threshold $\tau$}
	\end{minipage}
	\caption{The result of performance and fairness under dynamic data distribution and threshold}
	\label{3: dynamic data volume}
\end{figure*}

Fig.~\ref{3: dynamic data volume}(a) and Fig.~\ref{3: dynamic data volume}(d) present the experimental result under the dynamically uneven data distribution without overlap. All users start training with the same amount of data but will increase at different exponential rates and eventually form an extremely uneven distribution. 

First, for the \emph{Acc}, at the beginning, PoD, CFL and PBFT converge at the same rate, but inflexion points of convergence occur as data volume differentiates, as shown in Fig.~\ref{3: dynamic data volume}(a). This is because the threshold plays a role, and some nodes with a large amount of data are too late to participate in the settlement, thus significantly reducing the convergence speed. Subsequently, these big data nodes will not have the opportunity to participate in the settlement. Hence, the accuracy of the final training result is slightly lower than the traditional CFL, which gets a final test accuracy of 92.34\%, while PoD only reaches 90.34\%. Meanwhile, as discussed in section~\ref{1: data volume}, the differentiation of data volume will lead to system unfairness. Therefore, as the number of training epochs increases, user data volume differentiation and system fairness decrease, as shown in Fig.~\ref{3: dynamic data volume}(d).

Second, the attacks from Byzantine nodes can not significantly affect the system's performance but can result in significant unfairness in the dynamic environment. From Fig.~\ref{3: dynamic data volume}(a), it can be seen that the test accuracy of PoD-Byzantine is close to that of PoD without Byzantine nodes. This slight discrepancy can be interpreted as a loss of the Byzantine nodes' own data that does not participate in the computation and the lower threshold $\tau$ that honest nodes cannot participate in the settlement, and the attacking behaviour of Byzantine nodes does not affect the system at all. Moreover, from Fig.~\ref{3: dynamic data volume}(d), it can be seen that the system difference and max difference of PoD-Byzantine are larger than those of PoD without Byzantine nodes. In order to resist attacks of Byzantine nodes and guarantee the liveliness of the system, PoD-Byzantine sets a lower threshold $\tau$ than honest PoD that causes several honest training nodes to be unable to participate in epoch settlement, which results in significant unfairness. Moreover, the deposits of Byzantine nodes are also awarded to honest nodes participating in the epoch settlement.

In summary, the PoD consensus protocol had a close performance to conventional CFL before the huge divergence in data volume and had excellent fairness without Byzantine nodes. However, the performance and fairness of the system decrease obviously after the user data volume is greatly differentiated. Meanwhile, the PoD consensus protocol is Byzantine tolerant, and the tolerant ability is relevant to the voting threshold $\tau$. 

\subsubsection{Dynamic data volume with overlap}
Fig.~\ref{3: dynamic data volume}(b) and Fig.~\ref{3: dynamic data volume}(e) present the experimental result under the dynamically uneven data distribution with overlap. All users start training with the same amount of data without overlap but will increase at different exponential rates with 50\% overlap, and eventually form an extremely uneven distribution with a large amount of overlap. 

From the perspective of \emph{Acc}, the proposed PoD consensus protocol also outperforms the conventional CFL before huge user data volume differentiation, as shown in Fig.~\ref{3: dynamic data volume}(b). The main reason is that the novel contribution measurement system greatly supports the \emph{global aggregation} phase. User data contribution measurement system measures user aggregating factors from data volume and data uniqueness comprehensively, unlike the conventional CFL, which only considers the user data volume. Therefore, the PoD consensus can effectively eliminate the impact of data redundancy on system performance. However, when the amount of data is greatly differentiated and affected by the voting threshold $\tau$, the advantage of the PoD consensus protocol is not obvious.

Besides, for the \emph{Sys-diff} and \emph{Max-diff} of honest PoD shown by Fig.~\ref{3: dynamic data volume}(e), results similar to that under uniform data volume distribution as shown in Fig.~\ref{3: dynamic data volume}(d). As shown in the figure, the honest PoD also has similar system differences and maximum differences under the uniform data distribution. This result indicates that the fairness of the system is closely related to the distribution of data volume but not to the data overlap. Finally, Pod consensus can not defend against \emph{data redundancy attack}. System performance and fairness are significantly reduced.

To sum up, with the support of a user data contribution measurement system, data overlap does not significantly affect the performance and fairness of PoD. However, because the PoD consensus does not consider the quality of model updates, the system cannot resist \emph{data redundancy attack}.

\subsubsection{Dynamic threshold $\tau$}
From the above experimental results, we can obtain that there is a game between fairness (i.e., higher threshold $\tau$) and liveliness (i.e., lower threshold $\tau$) of the Byzantine system, and we need to look for a balance to maximize the performance of the system. To figure out how the voting threshold $\tau$ affects our solution, we also test the \emph{Acc}, \emph{Sys-diff} and \emph{Max-diff} under dynamic extremely uneven data distribution with overlap, and the experimental results are shown in Fig.~\ref{3: dynamic data volume}(c) and Fig.~\ref{3: dynamic data volume}(f).

% dynamic threshold
% dataset: dynamic, overlap, with/without Byzantine (3 line)
% analyze the influence of dynamic threshold
% Comparison with static threshold
From the perspective of \emph{Acc}, the proposed PoD consensus protocol also outperforms the conventional CFL before huge user data volume differentiation, as shown in Fig.~\ref{3: dynamic data volume}(c). The main reason is that the novel contribution measurement system greatly supports the \emph{global aggregation} phase. User data contribution measurement system measures user aggregating factors from data volume and data uniqueness comprehensively, unlike the conventional CFL, which only considers the user data volume. Therefore, the PoD consensus can effectively eliminate the impact of data redundancy on system performance. However, when the amount of data is greatly differentiated and affected by the voting threshold $\tau$, the advantage of the PoD consensus protocol is not obvious.

Besides, for the \emph{Sys-diff} and \emph{Max-diff} of honest PoD shown by Fig.~\ref{3: dynamic data volume}(f), results similar to that under uniform data volume distribution as shown in Fig.~\ref{3: dynamic data volume}(f). As shown in the figure, the honest PoD also has similar system differences and maximum differences under the uniform data distribution, and both differences are much less than 1. This result indicates that the system's fairness is closely related to the distribution of data volume but not to the data overlap.

In summary, the dynamic threshold can effectively reduce the inequity caused by uneven data distribution and improve the accuracy of the training result. But there will be temporary inequities.
\section{Related work} ~\label{sec: related work}
In this section, we first provide a brief introduction to the development of federated learning, and then summarize some related works about consensus algorithms in decentralized federated learning.

\subsection{Blockchain-enabled Federated Learning} ~\label{subsec: FL}
Konecn et al. proposed Federated Learning whose goal is to train a high-quality centralized model while training data remains distributed over a large number of clients~\cite{konevcny2018federated}. Subsequently, FL is applied in many scenarios like video analysis, information inspection, and classification, and credit card fraud detection and so on~\cite{chahoud2023feasibility,farooq2023ffm,chowdhury2023federated} while keeping personal data sensitivity safe. Besides, the theoretical studies of convergence~\cite{dinh2020federated,wei2022federated}, network latency~\cite{zhao2021system}, or malicious attacks~\cite{zhang2022fldetector,tolpegin2020data} on FL are also active fields.

Meanwhile, The centralized federated server has been challenged and questioned growly in these years. It is a natural thought that keeps the concept of server at a minimum or even avoids it completely. The study of~\cite{hegedHus2019gossip} assumed that the data remains at the edge devices, but it requires no aggregation server or any central component. Hu et al.~\cite{hu2019decentralized} proposed a segmented gossip approach, which fully utilizes node-to-node bandwidth and then can achieve a convergence efficiently.

Moreover, decentralization may be the most direct way to avoid the risks in centralized federated learning. Blockchain, a distributed ledger technique, can store historical operations and keep them tamper-resistant. With the aim of the blockchain, collaborative machine learning methods can eliminate the centralized server and improve security. It is reasonable to assume that the clients in FL might be malicious. Therefore, the local updates from all clients should be recorded under blockchain-based FL settings. Nguyen~\cite{nguyen2021federated} presented an overview of the fundamental concepts and explores the opportunities of FLchain in MEC networks which systematically analyzes the challenges and opportunities when Federated learning meets blockchain, and studies~\cite{kim2019blockchain,majeed2019flchain,podgorelec2020machine,qammar2023securing,lin2024refiner} proposed  blockchain-based federated learning architecture to solve these challenges, such as focus on convergence speed, stability, attacks and so on. These blockchain-based learning methods can effectively
record the nodes’ performance to reduce malicious attacks.
However, there are still several main challenges, such as consensus efficiency, model security, framework scalability and so on.

\subsection{Consensus for Blockchain-enabled FL} ~\label{subsec: consensus in FL}
With the development of blockchain-enabled federated learning, a series of new consensus algorithms have been proposed to support decentralized federated learning systems.

Bao et al.~\cite{bao2019flchain} proposed a public blockchain-based FL architecture, which provides trusty consensus based on nodes’ data amount and historical performance. Yuzheng et al.~\cite{2020_li2020blockchain} proposed a blockchain-based decentralized federated learning framework called BFLC with committee consensus which uses blockchain for the global model storage and the local model update exchange.
Zhikun et al.~\cite{2021_chen2021dacfl} proposed a new DFL implementation DACFL with a first-order dynamic average consensus FODAC method to track the average model in the absence of the PS. Xidi et al.~\cite{1-qu2021proof} proposed an energy-recycling consensus algorithm PoFL reinvest the energy wasted in PoW puzzles computing to federated learning problems. Wang et al.~\cite{wang2022platform} proposed a energy-recycling consensus mechanism named platform-free proof of federated learning (PF-PoFL) to leverages the computing power originally wasted in solving hard but meaningless PoW puzzles to conduct practical federated learning (FL) tasks. However, these works leave the Byzantine nodes untouched.

\section{Conclusions} ~\label{sec: conclusion}
Over-reliance on the central PS makes the federated learning possibly paralysed when the server breaks down. To alleviate this single-point failure in conventional FL, in this paper, we devise a novel decentralized federated learning framework coined as Proof-of-Data (i.e., PoD) consensus protocol to solve the consistency and liveliness problem in decentralized and open-access systems with Byzantine nodes. To confirm the feasibility of PoD, we also deliver a theoretical analysis on the premise of some assumptions, which offers a liveliness and safety guarantee of our solution. Besides, we design specific experiments on ImageNet under static and dynamic allocations and analyze the performance and fairness of the PoD. The results verify the effectiveness and fairness of PoD under various data distributions and declare that PoD can maintain outstanding performance and fairness in most cases.

\begin{acks}
This research was supported by the Singapore Ministry of Education (MOE) Academic Research Fund (AcRF) Tier 1 grant (Project ID: 22-SIS-SMU-048). Any opinions, findings and conclusions or recommendations expressed in this material are those of the author(s) and do not reflect the views of the Ministry of Education, Singapore.
\end{acks}

%\clearpage

\bibliographystyle{ACM-Reference-Format}
\bibliography{sample}

\end{document}